# Predecoherence: before Decoherence and Collapse

by Roland Omnès

Laboratoire de Physique Théorique[*]
Université de Paris-Sud, Bâtiment 210, 91405 Orsay cedex, France

*Abstract*:

Predecoherence, as its name indicates, is the same physical effect as decoherence, originating in the same interactions with an environment, injecting also incoherence and breaking unitarity. But whereas decoherence acts immediately after a measurement, predecoherence is acting long before. It is also a very strong effect and its main properties are established in this paper, including generation, transport, damping, and stationary level.

A mechanism for objectification, or wave function collapse, is also proposed as consisting in a perturbation by predecoherence of the intricacy between a measuring system and a measured one. The theory is made explicit on a special example and the quantitative results are found sensible.

---

# 1. Introduction

In the early nineteen thirties, a famous problem was pointed out in the foundations of quantum mechanics, particularly by Von Neumann, Bohr and Schrödinger [1]. Two distinct laws of dynamics seemed required for wave functions, one of them being Schrödinger's unitary evolution and the other one a random objectification, or wave function collapse, which generated a unique macroscopic datum in a measurement. When summing up this situation in a later paper [2], Wigner wrote in the introduction: The simplest way that one may try to reduce the two kinds of changes of the state vector to a single kind is to describe the whole process of measurement as an event in time, governed by the quantum-mechanical equations of motion."

This sentence would express perfectly the aim of the present work but, after writing it, Wigner proceeded to show that the consistency between collapse and quantum dynamics was impossible and thus confirmed earlier views by Von Neumann and Schrödinger [3, 4].

Wigner's proof supposed however an isolated measuring system, although one knew since a long time that strict isolation is impossible for a macroscopic system [5]. This question became some time after the topic of decoherence theory [6-10], which led to remarkable results: Far from being negligible, the interaction of a measuring apparatus with the environment has drastic consequences, particularly an extremely rapid damping of macroscopic interferences. The existence of decoherence has now been experimentally confirmed [11].

The collapse problem took then a new turn: One knew then that no realistic measuring device can be isolated and keep an inner unitary dynamics. Interactions with the environment are never negligible. The only exceptions occur in experimental models of measurement, for instance in quantum optics, where the measured system and the measuring one are both in a pure state and "measurement" amounts to a quantum jump [12]. But decoherence did not solve the problem of collapse since the quantum probabilities (or squared amplitudes) of different measurement channels are conserved under decoherence,

Strangely enough, not much work has been devoted to a rather obvious question [13], which is concerned with the action of interactions with the environment *before* a measurement and the action of decoherence. One may call this effect *predecoherence*. The interaction with the environment is permanent and exists already long before a measurement. As a matter of fact, it exists not only in a measuring device but in any macroscopic object such as a clock for instance. One may ask therefore which are the consequences of such a permanent action on the dynamics of a measuring apparatus, strong enough to destroy a unitary Schrödinger evolution in a very short time, through the disappearance of non-diagonal elements in a density matrix. One will also go much farther along this direction and ask a still more significant question: Could it be that predecoherence has something to do with collapse?

In this respect, one must mention as a precedent the work of Philip Pearle [14] (see also. [13]). He considered for instance a measurement in which the measured state is a superposition $|s\rangle = \sum c_j |j\rangle$ of measurable states $j$ with squared amplitudes $p_j = |c_j|^2$. He assumed that some process (which he attributed to hidden non-linearity) could produce Brownian fluctuations in the quantities $p_j$. The Brownian character of the process is characterized through the random fluctuations $\delta p_j$ of the squared amplitudes $p_j$ during a short time interval $\delta t$, such that

$$\langle \delta p_j \rangle = 0 \quad \text{and} \quad \langle \delta p_j \delta p_k \rangle = A_{jk} \delta t / \tau_c . \qquad (1.1)$$

The correlations $A_{jk}$ are constrained by simple conditions resulting from the conservation law $\sum \delta p_j = 0$. Pearle supposed that these coefficients depend only upon time and upon the instantaneous values of the $p_j$'s. The factor $\tau_c$ in (1.1) is meant only as a time scale, which could also be considered as a time scale of collapse, because of the following considerations and theorem: A Brownian motion must always bring some squared amplitude $p_k$ to vanish first, at one time or another. If the corresponding state of the apparatus cannot be regenerated, a second amplitude, and then a third one and so on will vanish, until a unique one, say $p_f$ survives. During the random process $p_f$ fluctuated, but it has become finally equal to 1, which means certainty in probability calculus. This final outcome can be therefore considered as an objectification, or collapse on Channel $j$ of the apparatus state.

The depth of this idea is shown by the value of the Brownian probability $P_f$ for getting the final result $p_f = 1$, which is

$$P_f = p_j(t=0) \text{ or } P_f = |c_f|^2. \tag{1.2}$$

This means that, if the assumptions of Pearle's version of the "gambler's ruin theorem" came to be justified, they would not only imply collapse, but also the validity of Born's probability rule linking the probability for getting a final datum to the square of the corresponding s amplitude in the initial state. This rule would become a consequence of the other quantum principles.

The research in the present paper rests on these earlier advances, through an extension of the decoherence conceptions to predecoherence and, as a consequence, a derivation from quantum mechanics of the assumptions underlying Pearle's theorem (with no appeal to nonlinear corrections to the Schrödinger dynamics).

Predecoherence is studied in Section 2. One finds that its main effect is an injection of incoherence into the quantum evolution of the apparatus, long before any measurement. Although much of the problem is new, one finds it manageable and obtains an explicit and coherent description for its continuous generation, its transport, damping, and the stationary level of perpetual incoherence resulting from it.

The relation of predecoherence with collapse, or rather with fluctuations in the channel probabilities, is considered in Section 3. The origin of fluctuations is attributed to a random continuous action of predecoherent disorder on the intricacy between the measured system and the measuring one. The undistinguishable character of atoms and the non-separable character of quantum mechanics enter in the existence of collapse. It turns out that, whereas decoherence is due to interactions with the environment occurring *after* the beginning of a measurement, collapse results from *previous* interactions, which occurred mostly long before measurement and were responsible for predecoherence. This difference between the actions of two effects, amounting to the same phenomenon, could explain why the program of decoherence stumbled on collapse: because one did not look at the meaningful order of events. Quantitative estimates for the rate of collapse are also made and yield sensible results.

Section 4 deals with some examples. In the case of the Schrödinger experiment [4], a problem is that there are different linear tracks [15] and they can compete. The problem is to understand how a unique track can come out from the competition. The Stern-Gerlach experiment and the measurement of two photons in an EPR pair, by space-like separated detectors [16, 17], are also considered and shown closely linked with the role of non-separability in the present theory.

Section 5 discusses briefly the meaning of incoherence at larger scale in the universe and states tentative conclusions.

**2. Predecoherence**

One will consider mainly in this paper a Geiger counter, denoted by $A$, which can detect charged particles. In the present section, there is no charged particle to be measured. The apparatus $A$ stands alone, though not perfectly isolated. It consists mainly of a solid box containing a gas of argon at atmospheric pressure. The nearby environment, denoted by $e$, is ordinary atmosphere.

The atmospheric molecules collide incessantly with the apparatus at a rate per unit time

$$\tau_p^{-1} = n_M v_M S, \qquad (2.1)$$

where $n_M$ and $v_M$ denote the space density and the average velocity of molecules and $S$ is the box area. This rate is extremely large and the collisions have therefore necessarily a strong influence on the quantum state of the apparatus. They produce particularly a high level of incoherence in the density matrix $\rho_A$, which will be called predecoherence in view of its common origin with decoherence and to mark on the contrary its durability.

Predecoherence is a very strong effect, which consists in an accumulation of many tiny effects over long periods. It is also a random effect and one will consider formally the series of collisions as a random process in which an event consists of a succession of collisions occurring at various places and at various times on the box.

*A formulation and a conjecture*

A random process implies in principle the notions of averages and fluctuations, which one considers now. One assumes for simplicity that the system $A$ and the surrounding atmosphere are at thermal equilibrium with the same temperature $T$. There is no measurement, or one considers the measuring device before measurement. One will show that the average quantum effects of predecoherence and the corresponding fluctuations can be expressed by a random density matrix $\rho_A(t)$, which consists of two parts:

$$\rho_A(t) = \langle \rho_A \rangle + \Omega(t). \qquad (2.2)$$

$\langle \rho_A \rangle$ describes the quantum properties of the system remaining insensitive to predecoherence, or averages over a sufficiently long time (which will be made precise). This average matrix is simply a thermal distribution

$$\langle \rho_A \rangle = Z^{-1} \exp(-H_A/T), \qquad (2.3)$$

where $Z$ is the partition function insuring the value 1 for the trace of $\langle \rho_A \rangle$. $\Omega(t)$ is a random self-adjoint matrix with zero trace, describing the fluctuations. Disregarding the possibility of zero eigenvalues, it can be split into two parts with respectively positive and negative eigenvalues so that $\Omega(t) = \Omega_+(t) - \Omega_-(t)$.

The relation (2.2) is somewhat obvious and will be justified, but real problems consist in finding explicit properties of $\Omega_\pm$. This will be also made, but remains somewhat conjectural, particularly concerning the traces of these matrices for which one will assume

$$Tr(\Omega_+) = Tr(\Omega_-) \approx 1. \tag{2.4}$$

The next four subsections will be devoted to the arguments leading to these conjectures, except for the last conjecture (2.4), which will be treated after a justification of (2.2) and (2.3).

(a) *Inadequacy of global wave functions*

A first argument is essential and goes against a tendency in measurement theory to attribute a central role to the wave functions of the apparatus and their unitary evolution. This is correct when the apparatus is isolated, but certainly not when interactions with the environment are taken into account.

To begin with, one will dispose of a few unessential features of the problem. An external collision affects initially some phonons in the box, which transmit the change of their state to the gas. There is no difficulty of principle in devising a theory of this transmission but one will avoid technicalities and consider that the gas directly feels the effect. The role of the box is then only to ascertain a definite atomic content of $A$ and thus allows describing $A$ as a well-defined though not isolated quantum system. For the same reason of simplicity, one will disregard the electronic and electric components of the detector, which can keep records when there is a measurement.

One considers then the academic case of an isolated detector in vacuum, undergoing a collision with a unique molecule $M$. The state of $M$ is taken as a wave packet $|G\rangle$, of which the location and average momentum insure that $M$ hits $A$ with certainty. The quantum probability $\langle G|G\rangle$ of the packet is denoted by $\varepsilon$ ($0 < \varepsilon \leq 1$). Let $|k\rangle$ denote the normalized eigenvectors (or wave functions) of the density matrix $\rho_A$ of $A$ and $p_k$ the corresponding eigenvalues. One also introduces an orthonormal basis of vectors $|u\rangle$ in the Hilbert space of $A$, including the eigenvectors $|k\rangle$ and possibly new states of $A$ arising the collision. The initial density matrix $\rho_A = \sum_k p_k |k\rangle\langle k|$ becomes a matrix $\rho'_A$ after taking a trace over the outgoing states of $M$, and one has

$$\rho'_A = \rho_A - \delta\rho_- + \delta\rho_+ \tag{2.5}$$

with
$$\delta\rho_- = \sum_k p_k |k\rangle\langle k| \left\{ \sum_{uq'} |T_{kG \to uq'}|^2 \right\} \tag{2.6}$$

and
$$\delta\rho_+ = \sum_{k,u'u'',q'} p_k T_{kG \to u'q'} T^*_{kG \to u''q'} |u'\rangle\langle u''|. \tag{2.7}$$

$T$ denotes the collision matrix, related to the $S$-matrix by $S = I - iT$. If $\tilde{G}(q)$ denotes the wave packet in momentum space, the $T$-matrix elements in these equations are given by

$$T_{kG \to uq'} = \sum_{u'} \tilde{G}(q) \langle uq'|T|kq\rangle \delta(E_k + E_q - E_u - E_{q'}).. \tag{2.8}$$

Strictly speaking, the delta function in energy is only manageable when the eigenstates $|k\rangle$ are also eigenvectors of the Hamiltonian $H_A$, which is the case when $\rho_A$ is in thermal equilibrium. The properties of $\delta\rho_+$ and $\delta\rho_-$ however, as they will be used, are valid more generally.

*Note*: A more explicit derivation of these expressions was given elsewhere [13]. One should mention however that introduction of arbitrary phase factors at a later stage in this reference was erroneous and some consequences would have to be revised (they are not used here).

The densities $\delta\rho_+$ and $\delta\rho_-$ in (2.7-8) are examples of contributions to $\Omega_+$ and $\Omega_-$ in (2.4). They are positive matrices (an obvious property for $\delta\rho_-$, which can be checked easily for $\delta\rho_+$). They have exactly equal traces:

$$Tr(\delta\rho_-) \equiv Tr(\delta\rho_+) = \varepsilon. \tag{2.9}$$

$\delta\rho_-$ has the same eigenvectors as $\rho_A$ and represents a decrement of the initial state, up to complete suppression when $\varepsilon = 1$, since the certainty of a collision implies

$$\sum_{uq'} |T_{kG \to uq'}|^2 = \varepsilon. \tag{2.10}$$

$\delta\rho_+$ represents similarly an increment in $\rho_A$ (or a reconstruction when $\varepsilon = 1$), through which $\rho_A$ is modified in a more or less drastic way.

When these equations are applied to the average matrix $\langle \rho_A \rangle$ in (2.2), every final state $|u\rangle$ in (2.6-8) is already occupied (*i.e.*, is a state $|k'\rangle$). For elastic or quasi-elastic collisions, one finds then that the diagonal parts of $\delta\rho_-$ and $\delta\rho_+$ are practically equal so that their addition to $\Omega$ is essentially the non-diagonal part of $\delta\rho_+$. Such a compensation does not hold however for the action of *M* on an already present $\Omega$ in (2.2).

The contribution of $\langle \rho_A \rangle$ when $\varepsilon$ is small is also instructive. If one applied perturbation theory to evaluate the eigenvectors of the perturbed density matrix $\rho'_A$, one would get in first order

$$\delta|k\rangle = \sum_{k'} |k'\rangle\langle k'|(\delta\rho_+ - \delta\rho_-)|k\rangle/(p_k - p_{k'}). \tag{2.11}$$

But the eigenvalues $p_k$ are so close that in spite of the smallness of $\varepsilon$, this series cannot make sense. This means essentially that, because of the extreme proximity of eigenvalues in the spectrum of $\langle \rho_A \rangle$ and degeneracy, the wave functions are extremely sensitive to external collisions.

To appreciate this point better, one may think of the high complexity (in an algorithmic sense) of the wave functions for the macroscopic system *A*. The effect of an external collision is tiny, in so far that *the set* of wave functions looks essentially the same after a collision as it looked before. The change has therefore no consequence on the average of a realistic physical quantity (although is could be significant for a mathematical observable). The individual wave functions in the set are on the contrary strongly modified and their unitary evolution is broken. This vulnerability will be considered as a warning concerning the relevance of global wave functions.

(b) *Transport of predecoherence*

A collision of an external molecule occurs at some time and in some region on the solid envelope of the apparatus *A*. Its effect on the state of *A* is initially located near that region and expands from there to the bulk of *A* through atomic collisions. Such a behavior can be at least expected in a phenomenological approach, and one will now look at it.

In statistical physics, locality is usually described in terms of a coverage of *A* by a set of small though macroscopic Gibbs cells $\beta$. The density matrix $\rho_\beta$ of a cell is defined as a partial trace of $\rho_A$ over the values of quantum fields outside $\beta$ (in the present example, these fields are defined in a second-quantized formalism [18]). An approximate factorization of $\rho_A$,

$$\bar{\rho}_A \approx \prod_\beta^\otimes \bar{\rho}_\beta, \qquad (2.12)$$

can then express locality after an average over a convenient short time, denoted by an upper bar. Although the exact meaning of the approximation in (2.13) and its limitations are not obviously clear, this expression is very useful in practice and particularly significant for an introduction of entropy [19].

When this approach is adapted to the present problem, it goes as follows: One introduces a Gibbs cell $\beta$ containing the atoms (or phonons), initially affected by the collision. One also introduces the complement $C\beta$ of $\beta$ in *A*, which is a large region involving most of *A*. The essential point in locality, at least in the initial stage, is that the state of $C\beta$ receives no influence from the collision, suggesting a factorization

$$\rho_A \approx \rho_{C\beta} \otimes \rho_\beta. \qquad (2.13)$$

Like (2.13), this factorization is tricky from a fundamental standpoint, mainly because it does not use a time average. One will assume however that (2.13) is significant for an account of a predecoherent perturbation. To obtain this account, one introduces the perturbations $\delta\rho_{\beta+}$ and $\delta\rho_{\beta-}$ as they become in $\rho_\beta$, and writes down

$$\delta\rho_A \approx \rho_{C\beta} \otimes (\delta\rho_{\beta+} - \delta\rho_{\beta-}). \qquad (2.14)$$

This equation does not assume the validity of (2.13) (with its absence of an average on time). It relies only on a supposed lack of effect of the microscopic perturbation outside of the macroscopic region $\beta$.

Since the motion of atoms and atomic collisions control the internal dynamics of the gas, an initial local perturbation is transported into the bulk of *A*. If there is no strong damping, the cell $\beta$ containing the fluctuation grows, its boundary inflates, while the outside region $C\beta$ shrinks eventually until $\beta$ fills up the whole of *A*. One will try now to get something more about this transport process.

(c) *Predecoherence waves*

To investigate the transport of predecoherence, it will be convenient to consider a case in which the environment involves also electromagnetic radiation: The box enclosing *A* does not only interact with the atmosphere, but also with a flash of light illuminating *A* on one of its faces, denoted by *F*, during a short time. The matter in *F* reflects the light or diffuses it and one will not enter into a discussion of the corresponding physical mechanism. The light beam is much wider than *F* and its photon states are plane waves parallel to *F*. There is then diffraction in addition to reflection.

One disregards the intensity of the effect (*i.e.*, the value of $\varepsilon$). There are again two perturbations $\delta\rho_+$ and $\delta\rho_-$, initially located along $F$. A new feature is that they are not only distinguished by their sign and their origin, but also by intricacy: The eigenvectors of $\delta\rho_+$ are intricate with reflected photon states and $\delta\rho_-$ with diffracted ones. This is obvious for $\delta\rho_+$ but less trivial for $\delta\rho_-$ and one must go back to scattering theory as in [13] to prove it. For brevity, one mentions only that (2.10), which resulted from unitarity of the *S*-matrix, is also now an expression of Babinet's principle. As before, one is interested in the transport of the perturbations in the gas.

When two atoms *a* and *a'* collide and the state of *a* belongs to $\delta\rho_+$ (so that it is intricate with reflected light) while *a'* is not intricate with light, the two atoms in the pair *aa'* become intricate with reflected light after the collision. The same is true when the state of *a* belongs to $\delta\rho_-$ and is intricate with diffracted light. There is therefore a transport of the predecoherent perturbations, due to atomic collisions, which can expand a priori with a velocity of the order of the thermal velocity of atoms.

To show that this is the case, one must consider the populations of intricate and non-intricate atoms. Let $f(x,t)dV$ denote the average number of atoms having intricacy with reflected light in a small volume $dV$, and $f_0 dV$ the average total number of atoms in this volume, either intricate with reflected light or not. If the motion of atoms is considered as Brownian through their mutual collisions and if the mean free path $\lambda$ and mean free time $\tau$ of an atom are taken as units of length and time, the evolution of $f(x,t)$ is given by

$$\partial f / \partial t = (1/2)\Delta f + f(f_0 - f) \qquad (2.15)$$

The first term in the right-hand side represents the Brownian motion of intricate atoms colliding together, which is dominant in a region where most atoms have already become intricate (one took the random walk value 1/2 for the diffusion coefficient). The second term represents the local increase in the population of intricate atoms because of their collisions with non-intricate ones. The factor $1 - f/f_0$ represents the fraction of non-intricate atoms, witch have a space density $f_0 - f$.

The nonlinearity of (2.16) suggests the possibility of a wave motion with a moving front, which would separate a region, bounded by $F$ and the front and containing intricate atoms, from a region not yet reached by the wave, where no atom is intricate and $f(x,t) = 0$. When looking at this possibility, one notices first that the relative number of intricate atoms $f/f_0$ must tend to zero near the front. The equation (2.6) is not valid however near the front where $f(x,t)$ is small. There are only few intricate atoms and their collisions and motion are governed by discrete random walk rather than by the first diffusive term in the right-hand side of (2.5).

One considers therefore a one-dimensional discrete random walk model where representative intricate atoms are located on planes, parallel to the front and separated by a distance $3^{-1/2}\lambda$ in the direction *x*, normal to the front (the square root of 3 takes into account that the collisions are three-dimensional). A collision of an intricate atom occurs in practice only with non-intricate ones in the front region. Moreover, after a collision, the two atoms are intricate and not distinguishable (because of Bose-Einstein or Fermi-Dirac symmetry). Except for some cases where the two atoms move in the same directions and one of them catches the other one. One atom moves on average with velocity $3^{-1/2}v$ in the *x*-direction and the other atom in the opposite direction with an equal average velocity

In the discrete random walk model, one may consider that every intricate atom is duplicated after a time $\tau$, one copy making one step in the *x*-direction and the other copy in the opposite direction. Assuming a stationary shape of the front, this means that it moves with velocity $v$, *i.e.* 1 in the present units

Equation (2.16) becomes then

$$\partial g / \partial t = \frac{1}{2} \partial^2 g / \partial x^2 + g(1-g) \tag{2.16}$$

for the function $g = f / f_0$. The boundary conditions are $g = 0$ on the moving boundary $x = t$ and $g = 1$ at the position $x = 0$ of $F$.

The same results are valid for $\delta\rho_-$ because of its intricacy with refracted photon states. The argument of intricacy is not valid apparently in the case of collisions by external molecules and one is left to suppose that predecoherent waves, if they exist, result from a combination of the evolution equation for local perturbations of the Wigner function of $\rho_A$ with account of its symmetries. This is a long shot and, with due caution, one will only conclude that predecoherence moves (rather than diffusing), which is enough to understand something of it.

(d) *The question of perturbations damping*

The question of possible cancellations between positive and negative perturbations is of obvious importance. Analogous examples exist in Boltzmann kinetics and one of them is especially significant [21]: It deals with an experiment in which a powerful laser flash is focalized on a gas. Energetic particles are produced at the focus and they collide with gas molecules, producing a first generation of perturbations with increased velocities. These more rapid molecules collide then with slower ones, with which they share some of the excessive velocity, producing thus a second generation, and so on. In the standard Boltzmann distribution, two terms appear then. One of them is positive and corresponds to the arrival of molecules in a previously empty region of phase space. The other one is negative and describes the corresponding loss in the occupied region. It has been shown that, after some time, the loss in excess of velocities brings back the Boltzmann distribution to thermal equilibrium and perturbations vanish.

This example is difficult to transpose however because it deals with the Boltzmann reduction to one particle of a classical distribution of many particles, whereas one is interested presently in the whole quantum state of a very large number of particles. The underlying concepts need an elucidation, which can be attempted as follows:

One will say that two density matrices $\rho$ and $\rho'$ are equivalent in a phenomenological sense – or more simply that they look alike – when they yield approximately the same local values for few-particle states, by which one means the same few-particles density matrices (few meaning for instance 1 to 3 if three-atoms scattering were taken into account). In that sense, it is reasonable to assume that $\rho_A(t)$ in (2.2) looks the same as $\langle \rho_A \rangle$.

These two matrices are very different however and one needs an expression of this difference. A well-known one is the matrix distance, namely: One can define a real scalar product between these self-adjoint matrices as

$$\{\rho,\rho'\} = Tr(\rho\rho'). \tag{2.17}$$

A distance between $\rho$ and $\rho'$ is associated with this scalar product and is given by

$$d(\rho,\rho') = \left(Tr[(\rho-\rho')^2]\right)^{1/2}. \tag{2.18}$$

A convenient measure for the similarity of measure of their similarity of $\rho$ and $\rho'$ is then given by the number

$$K = Tr(\rho\rho')/[Tr(\rho^2)Tr(\rho'^2)]^{1/2}. \tag{2.19}$$

It is clear that $K = 1$ is a necessary and sufficient condition for equality of $\rho$ and $\rho'$. One will say on the other hand that two matrices $\rho$ and $\rho'$, which look alike, are foreign to each other when $K$ vanishes. This notion is convenient for distinguishing two density matrices yielding the same macroscopic and microscopic statistics, and however very different in their eigenfunctions, *i.e.*, at a mathematical or fundamental level. A drawback is of course the recourse to the traces squares and products of density matrices, which are not much manageable. One can make often however educated guesses about the extreme cases when $K = 0$ or $1$, which can help much in making sensible conjectures.

*Return to the conjecture* (2.2)

The previous subsections reviewed the main remarks leading to the conjecture (2.2), which one will now formulate more properly. One recalls that the average defining $\langle\rho_A\rangle$ is meant in the framework of a random process, in which each individual event consists in a series of external collisions at different locations on the box and at different times. Two distinct events differ by the number, the locations and times of their respective series of collisions. This random process is essential in the present approach, particularly for the existence and randomness of fluctuations in the channel probabilities when a measurement occurs, as in section 3. Deep questions concerning the interpretation of quantum mechanics stand obviously behind this, particularly the quantum behavior of the environment, but they will be delayed until their examination in a wider framework in Section 5.

The average $\langle\rho_A\rangle$ accounts then for the properties of $\rho_A(t)$, common to all the random events. Since the random process of external collisions acts permanently and is invariant under a time translation, the average $\langle\rho_A\rangle$ must be stationary and can involve only the average value of the total energy, which is the only observable remaining insensitive to external collisions in the present case. The expression (2.3) is then obviously the only possible one for it.

The matrix $\Omega(t)$ represents the deviation from $\langle\rho_A\rangle$ of the density matrix $\rho_A(t)$ of $A$ at some time $t$ when $A$ underwent a specific series of external collisions, the matrices $\Omega_\pm(t)$ being then well defined as well as its two parts with respective positive and negative eigenvalues.

The time evolution of the density matrix $\rho_A(t)$ results from internal and external dynamical effects. The internal ones, including the transport of predecoherence in $A$, are governed by the Hamiltonian $H_A$ of $A$. The external effects consist of the arrival of external molecules on the box and one can thus write

$$i\partial\rho_A/\partial t = [H_A,\rho_A] + S(t), \tag{2.20}$$

where $S$ is a random source of predecoherence, directly due to the external collisions, and is located on the box. As a matter of fact, $S(t)$ depends on $\rho_A(t)$, as shown in (2.6,7), so that (2.17) is nonlinear, but one does not need an explicit account of this property.

Integrating this equation from a previous time $t'$ to time $t$, one gets

$$\rho_A(t) = U(t-t')\rho_A(t')U^\dagger(t-t')$$
$$+ \int_{t'}^{t} dt'' U(t-t'')S(t'')U^\dagger(t-t''), \quad (2.21)$$

with $U(t) = \exp(-iH_A t)$.

One can use this equation to get a better understanding of the matrices $\Omega_\pm$ and explain the conjecture (2.4). This is the trickiest part of the discussion, because it relies on the notion of similar though foreign matrices, introduced in Subsection (d), which sounds obvious but is however unproved.

One begins by introducing an ideal case where the system $A$ is subjected to predecoherence until a time $t'$ after which $A$ is isolated from the environment. One has then $S(t) = 0$ for $t > t'$ and the density matrix becomes

$$\rho_{A,iso}(t) = U(t-t')\rho_A(t')U^\dagger(t-t'). \quad (2.22)$$

If $t - t'$ is somewhat larger than the time during which a predecoherent wave crosses $A$, one can expect that thermal equilibrium has occurred at time $t$. This does not mean however that $\rho_{A,iso}(t)$ is exactly equal to the equilibrium matrix $\langle \rho_A \rangle$, but only that they are similar in the previous sense: they yield only the same practical predictions. On the other hand, there exist certainly some mathematical observables (conveniently constructed operators) for which they yield very different average values. Said otherwise, their sets of eigenfunctions look similar but the eigenfunctions themselves are very different in detail, mainly because of the algorithmic complexity of the Schrödinger equation for macroscopic systems. One will consider therefore that $\rho_{A,iso}(t)$ and $\langle \rho_A \rangle$ are foreign to each other. One thus gets immediately a formal expression for $\Omega(t)$, which is

$$\Omega(t) = \rho_A(t) - \rho_{A,iso}(t). \quad (2.23)$$

But $\rho_A(t)$ and $\rho_{A,iso}(t)$ are again similar and foreign to each other (their sets of wave functions look the same but are very different in detail). One can then assimilate the positive and negative parts of $\Omega(t)$ to

$$\Omega_+(t) \cong \rho_A(t), \qquad \Omega_-(t) \cong \rho_{A,iso}(t), \quad (2.24)$$

where the sign $\cong$ (congruent to) between two matrices means similarity in their sets of eigenfunctions with a small value of the correlation $K$ in (2.19). An approximation in this congruence is due partly to arbitrariness in the choice of the time $t'$, but also and jointly to our ignorance concerning the dynamics of predecoherence damping, which is necessary to reach the following limit, resulting from (2.24):

$$Tr[\Omega_+(t)] \cong Tr[\Omega_-(t)] \approx 1. \quad (2.25)$$

The existence of damping is necessary for the validity of these relations. If there were predecoherent waves as in Subsection (c), and if they did not cancel partly by mutual deflation and reconstruction, their number would be very large, of order $N = L/v\tau_p$, if $L$ denotes a typical size of the apparatus. In view of (2.9), with $\varepsilon = 1$ for the sum of collisions occurring during a time $\tau_p$, one would get the absurdly large value $N$ for the trace of $\Omega_\pm$. On the other hand the congruence (2.24), although relying on the unsure notion of similar and foreign matrices, is more credible and will be used in the next section.

Perhaps the best argument for formulas such as (2.25) and for their simplicity is intuitive: One knows the great intensity of decoherence, since it can destroy the quantum correlations in non-diagonal matrices showing macroscopic interferences in an extremely short time. On the other hand, predecoherence acts not as a flash like decoherence, but always with the same great strength. Because of its destructive and creative characters, it must bring itself its own limitation, in a permanent way. This argument does not yield trivially the result (2.25), but only that some actual limit of that sort should exist. It would be nice to find a more cogent and elegant proof for it than the present one.

Finally, one should stress that these results rely heavily on the very large value of the predecoherence rate (2.1). This rate is practically zero for a microscopic system and certainly negligible for a mesoscopic system, since it requires a sufficient size. In the mesoscopic case, predecoherence can exist but remains a perturbation.

*Addendum*

One adds here a comment on the generalization of (2.2), because a special case will be need in the next section. The present approach holds for many other macroscopic systems. They can be also systems with organization, such as for instance a mechanical watch: There are also external collisions on the watch and one might be interested eventually in its quantum state. The main properties having no influence from predecoherence –or rather from external injection of incoherence– would be then the positions and velocities of the wheelworks. They depend on time and obey classical dynamics, but this classical motion results fundamentally from quantum dynamics and the corresponding properties can be expressed by microlocal projection operators, accounting simultaneously for positions and momenta [20].

A general expression of $\langle \rho_A \rangle$ can be obtained in these cases: One denotes by $A_n$ the observables with average value insensitive to the environment (this was the case for the total energy in $\langle \rho_A \rangle$). One denotes also by $P(t)$ the product of all the projection operators expressing the macroscopic properties of the system at time $t$. The general form of (2.2) is then:

$$\rho(t) = P(t)\exp(-\sum_n \lambda_n A_n)P(t) + \Omega(t). \qquad (2.26)$$

The quantities $\lambda_n$ are Lagrange parameters, which insure correct average values for the observables $A_n$. This expression insures also the condition $Tr(<\rho>) = 1$ when one includes the identity operator among the operators $A_n$. Generally, the average values, the Lagrange parameters and the classically meaningful projections depend on time. Slow effects, such as for instance a difference in temperatures between $A$ and $e$, could be described in principle by (2.26).

Finally, one may notice that (2.26) coincides in practice with the density matrix resulting from information theory [22], although there is much difference in the underlying concepts.

*A final overview*

One will use the results of the present section in the next one to examine the problem of objectification. This is a deep problem, which requires a clear statement of what is held for sure (or explicitly assumed) in its formulation. Since the present results will be used, it seems appropriate to look back now at the foundations of the whole approach.

First of all, the principles of quantum mechanics are taken as granted and used as absolute. Secondly, since one must use the notion of environment and there is no way to define it exactly, every physical system must be envisioned as a part belonging to a whole, which is the universe. This is not explicitly stated usually among the principles of quantum mechanics, but one will assume as a principle that the universe is a quantum system with a sufficient approximation and, moreover, any local quantum system is a subsystem of this largest one.

This statement has consequences, for instance in the definition of the state of a local system $S$ by a density matrix. There is no obvious and general way to substantiate this definition, if not through a trace over the values of quantum fields outside $S$. It requires however that the state of the largest system exists, at least formally, so that one can normalize it and, from there, deduce that the density matrix of $S$ is also normalized (with trace 1). This approach to normalization might look unnecessary, but it has a central consequence: namely that, whatever approximations one can be led to, the agreement with normalization must be absolute.

When the existence of an environment (*i.e.*, of the universe) cannot be legitimately neglected, the density matrix $\rho_S$ of a system $S$ can be supposed defined at a sufficiently sharp time. One does not know it however and different theoretical constructions can lead to different representations for $\rho_S$ with equivalent empirical consequences. Nonetheless, the absolute character of the quantum principles requires that the difference between the explicit mathematical expressions of these representations should be acknowledged. This requirement was used here as the basis for distinguishing between conguent representations.

Since one can assume in the present case the empirical representation $\rho_A = \langle \rho_A \rangle \pm \Omega_\pm$, and since the exact $\rho_A$ is only defined at a sufficiently sharp time, the predecoherent matrices $\Omega_\pm$ are also defined within this bound on time. When time evolves, $\Omega_\pm$ are nourished by the environment, but they act also on $\langle \rho_A \rangle$ as well as they act on each other in view of their damping. This interaction between different formal components of the predecoherent representation of $\rho_A$ will become central in the next section.

**3. A model of collapse**

*A model of measurement*

One considers a measurement in which an external source emits an alpha particle (denoted simply by $\alpha$) in a pure superposed state $|s\rangle\langle s|$ where

$$|s\rangle = c_1|1\rangle + c_2|2\rangle. \tag{3.1}$$

The two states $|1\rangle$ and $|2\rangle$ of the particle $\alpha$ travel along different straight lines. One will distinguish two cases: In Case I, $\alpha$ crosses the previous Geiger counter $A$ in state $|1\rangle$, but does not cross it in state $|2\rangle$. One will then say then that Channel 2 is mute. In Case II, $\alpha$ crosses $A$ in both states $|1\rangle$ and $|2\rangle$, but the two trajectories are sufficiently distant to yield clearly distinct tracks.

The particle slows down in the detector and generates a track consisting mostly of excited atoms an also a smaller number of ions and free electrons. An electric field slightly modifies the trajectories, but its main effect consists in generating a cascade of secondary excitations and ionizations, which produce a record after reaching a macroscopic level.

When $\alpha$ and $A$ interact, their states become entangled and the density matrix of the system $A + \alpha$ becomes

$$\rho_{A+\alpha} = \sum_{j,j''} \{|j\rangle\langle j'|\} \otimes \rho_{Ajj'}, \tag{3.2}$$

where $j$ and $j'$ are equal to 1 or 2. One has then

$$Tr(\rho_{Ajj}) = p_j \equiv |c_j|^2. \tag{3.3}$$

One is interested in eventual fluctuations in the traces ($p_1, p_2$) of the two "diagonal" matrices $\rho_{A11}$ and $\rho_{A22}$, which would occur after track formation. The "non-diagonal" matrices $\rho_{A12}$ and $\rho_{A21}$ are rapidly damped by decoherence, but this point is unessential (if there is collapse, macroscopic interferences disappear anyway).

Fluctuations in ($p_1, p_2$) will be shown a consequence of predecoherence and, more precisely, of predecoherence already present before the measurement. Using the expressions (2.2-4), one introduces therefore predecoherent average density matrices $\langle \rho_{A11} \rangle$ and $\langle \rho_{A22} \rangle$, and also the random matrices $\Omega_+$ and $\Omega_-$. The average non-diagonal matrices $\langle \rho_{A12} \rangle$ and $\langle \rho_{A21} \rangle$ vanish, contrary to their actual realizations $\rho_{A12}$ and $\rho_{A21}$ whose traces only vanish under decoherence. (This is because the matrix elements of $\rho_{A12}$ and $\rho_{A21}$ are scalar products of eigenvectors of $\rho_{A11}$ and $\rho_{A22}$, which are extremely sensitive, so that their products vanish on average). In any case, one will not have to deal with them.

*Slowing-down and intricacy*

The slowing down process obeys Schrödinger's dynamics, which predicts that $\alpha$ follows a definite trajectory, which one assimilates to a straight line, and one introduces a distance $x$ along this line. The Bethe-Bloch equation describing the slowing down process predicts the average number density $n(x)$ of excited and ionized atoms along the track, while atomic physics predicts the quantum average of excitation energy [23]. One will denote by $N(x)$ the observable with quantum average $n(x)$, which can be identified with an operator $a^\dagger(x)a(x)$ in terms of quantum fields $a(x)$ describing excited atoms [18]. The "density effect correction" in the Bethe-Bloch equation is negligible in the present case and the atoms can be considered as independent in the slowing-down process. As a consequence, the predictions are identical for the average density matrix $\langle \rho_{A11} \rangle$ and for the predecoherent one $\rho_{A11}$.

One will consider for simplicity that the process is quick enough to make the $\alpha$-$A$ interaction almost instantaneous and occurring at some time 0. One will also neglect the difference between excitation and ionization and consider all the interacting atoms as excited with the same energy. Moreover, one will concentrate attention on the effects occurring soon after time 0 and pay no special attention to free electrons and the production of cascades.

Because predecoherence breaks down global wave functions and the unitary evolution they would have in ideally isolated systems, it brings emphasis on locality and transport. One therefore introduces a macroscopic region $R$ in which the track is contained (the "reactive region" in [13]), and also a region $B$ (the "background"), which is the complement of $R$ in $A$. This is enough in Case I where Channel 2 is mute, but there are two distinct reactive regions in Case II and $B$ is the complement of their union. For simplicity, most of the following discussion will be made for Case I.

The initial detection effects take place in $R$ whereas $B$ is insensitive to them, so that one can use an approximate factorization just after time zero, in which

$$\langle \rho_{A+\alpha} \rangle \approx \left\{ \sum_j |j\rangle\langle j| \otimes \langle \rho_{Rjj} \rangle \right\} \otimes \rho_B. \qquad (3.4)$$

.
Eq.(2.3) yields in Channel 1:

$$\langle \rho_{R11} \rangle = \exp(-\lambda_1 I - H_R/T - \mu_1 N). \qquad (3.5)$$

with an abbreviated notation where

$$\mu_1 N = \int \mu_1(x) N(x) dx. \qquad (3.6)$$

These equations rely again on the fact that an average matrix can express only properties that are insensitive to predecoherence. The Lagrange parameters insure respectively a trace of $\langle \rho_{A11} \rangle$ equal to $p_1$ (for $\lambda_1$), the thermal average energy in the medium (for $T$) and the average values $n(x)$ of the numbers $N(x)$ (for $\mu_1(x)$). One did not write explicitly the contribution of excitation to the Hamiltonian, because it derives from $n(x)$ when average excitation energy is used.

One can also use (3.4) and (3.5) to write down

$$(|1\rangle\langle 1|) \otimes \langle \rho_{R11} \rangle = (|1\rangle\langle 1|) \exp\{-|1\rangle\langle 1| \otimes (\lambda_1 I + \mu_1 N) - (1/T) I_\alpha \otimes H_R\}, \qquad (3.7)$$

where $I_\alpha = |1\rangle\langle 1| + |2\rangle\langle 2|$ is the identity operator in the Hilbert space of $\alpha$. This equation shows clearly the locality of excitation and, moreover, the sum in the exponent is analogous to a factorization and shows something more: The quantum state of the whole region $R$ is not only factorized out from $\langle \rho_A \rangle$ as it was in (3.4), but there is a much finer separation between essential effects and inert ones, since the excited atoms in $R$ are singled out and the crowd of other atoms is left out, including those in $R$ that did not become intricate with $|1\rangle$.

Intricacy and entanglement become much thus more elaborate in a predecoherent system than they are in an ideally isolated one where the state is rigidly frozen into global wave functions. There is now an individuation of intricacy, which is initially restricted to the excited states, similar to excitons in solid-state physics [24]. This implies a departure from the usual interpretation of entanglement in measurement theory: The state $|1\rangle$ of the particle $\alpha$ *is* not directly intricate with all the atoms in the average matrix $\langle \rho_{R11} \rangle$, but only with excitons. Moreover, although this restriction to excitons is initially valid, there is soon after a growth and transport of intricacy, which selects new non-excited atoms for intricacy out of a crowd of spectators.

One encountered already this kind of effect in Section 2 with the growth and transport of intricacy in a predecoherent wave resulting from illumination. When an atom $a$, intricate with $|1\rangle$, collides with a non-intricate atom $a'$, both atoms are intricate with $|1\rangle$ in the final state. The number of intricate atomic states grows therefore. There is however no change in the channel probabilities ($p_1$, $p_2$), as long as everything occurs in $\langle \rho_{A+m} \rangle$, because the Hamiltonian $H_A$ governs the process unitarily. As shown in the previous section, there is a local growth in intricacy and also an expansion of the regions of intricacy, both of which start from the track. The boundary of the intricate region moves at the thermal velocity of atoms, while the local increase in the number density of intricate atoms is governed by Eq. (2.15). The local number of intricate atoms increases therefore initially like $\exp(t/\tau)$, where $\tau$ is the mean free time of atoms, and this increase saturates finally when all the local atoms have become intricate. It will be convenient from there on to coin a name for atoms intricate with $|1\rangle$, either excited or not, and one will call them "intricons", on the model of "excitons".

Nothing of that kind occurs on the other hand in the mute channel 2 and $\langle \rho_{A22} \rangle$ remains given by the simple form (2.3), except for its probability $p_2$. The necessary extension to two channels in Case II will be considered as trivial.

*Histories of collapse*

One will now anticipate on the existence of fluctuations in $p_1$ and $p_2$, which would result from the predecoherent matrices $\Omega_\pm$. The purpose is to identify the relevant theoretical concepts. One will mainly concentrate on Case I where there is a mute channel, because it is especially delicate and will be shown in Section 4 to play central part in some paradigms of measurement.

One builds up a history of collapse, starting at time $0_-$, just before the $\alpha$-$A$ interaction. The average matrix $\langle \rho_A \rangle$ is still given at that time by (2.3). One also assumes that, because of their saturation resulting from anterior mutual interactions and expressed by the equations (2.4), the two matrices $\Omega_+$ and $\Omega_-$ are independent.

The $\alpha$-$A$ interaction occurs at time 0. It would be easy to account for the duration of this interaction and its development in space, but this is unessential and one will not enter in these aspects. A new average matrix $\langle \rho_{A+\alpha} \rangle$ occurs at time $0_+$, immediately after the $\alpha$-$A$ interaction, and is written as

$$\langle \rho_{A+\alpha} \rangle = \sum_j \{|j\rangle\langle j|\} \otimes \langle \rho_{Ajj} \rangle. \tag{3.8}$$

This is not an abstract quantity but the exact outcome of the interaction between $\langle \rho_A \rangle$ and $|s\rangle\langle s|$, deprived of its non-diagonal terms. Actually, it depends much on preparation, through the exact wave functions of the incoming states $|j\rangle$.

After a short time $\delta t$, the states $\Omega_\pm$ have begun to act and to produce fluctuations in $p_1$ and $p_2$. There have been many interactions of $\Omega_+$ and of $\Omega_-$ with individual intricons and the resulting effects on the channel probabilities mostly cancelled, but fluctuations remain because of the independence of $\Omega_+$, $\Omega_-$, and their independence from the states of intricons. The outcome is a random change $\delta p_1$ in $p_1$ and $\delta p_2 = -\delta p_1$ in $p_2$, positive or negative.

To follow the development of the process, one represents these fluctuating effects by the introduction of two average density matrices in place of the unique one $\langle \rho_{A+\alpha} \rangle$ at time $0_+$. One might denote them with indices and a time reference, for instance $\langle \rho_{A+\alpha}^{(+)}(\delta t) \rangle$ and $\langle \rho_{A+\alpha}^{(-)}(\delta t) \rangle$, but the notation would become rapidly heavy and one will just say that two copies of $\langle \rho_{A+\alpha} \rangle$ heve been generated at time $\delta t$. In the first one, $p_1$ changed by a positive quantity $\Delta p_1$, defined by

$$(\Delta p_1)^2 = \langle (\delta p_1)^2 \rangle, \tag{3.9}$$

where the average in the right-hand side results from the many $\delta p_1$ from interactions of intricons with $\Omega_\pm$. The change in the value $p_2$ for the trace of this matrix is negative and equal to $-\Delta p_1$. In the second matrix, $p_1$ changed by $-\Delta p_1$ and $p_2$ by $+\Delta p_1 00$.

More fluctuations occur during the time interval $[\delta t, 2\delta t]$. Each matrix at time $\delta t$ gives rise to two copies with respectively positive an negative $\Delta p_1$, generally different from the previous $\Delta p_1$, so that one gets four copies according to the signs of fluctuations during each time interval $\delta t$. Clearly, at time $n\delta t$, $2^n$ different copies have been generated and each one of them is associated with a definite random walk in a two-dimensional space with coordinates $p_1, p_2$ (actually a one-dimensional space since $p_1 + p_2$ remains equal to 1). This is practically identical with Pearle's model and should imply therefore collapse with final probabilities for the outcome agreeing with Born's rule.

The external collisions occurring after time 0 do not contribute to this outcome, whereas they are responsible for decoherence in the non-diagonal elements of the matrix $\rho_{A+\alpha}$ resulting from interaction of $\alpha$ in the sate $|s\rangle\langle s|$ with the "actual" state $\rho_A$, given by (2.2). This lack of influence of decohering collisions is due to the slow transport of their effects, which only affect a margin of the $A$-boundary with width $v\tau_c$ if $\tau_c$ is the (very short) time of collapse. The intricons in region $R$ are generally far from this boundary and their interactions with atomic states in $\Omega_\pm$ are local. Fluctuations in probabilities are therefore unaffected by later effects from external molecules.

One should stress that this conceptual framework of collapse histories relies explicitly and necessarily on a "real" randomness of predecoherence, or of the matrices $\Omega_\pm$. This point could have far-fetched consequences: If there exists for instance a wave function (or a quantum state) of the universe, $\rho_A$ is perfectly well-defined at time 0 as a partial trace of this state and involves apparently no internal randomness. One will come back to this question in Section 6 and leave it aside presently.

*Fluctuations in probabilities*

The random predecoherent matrices $\Omega_\pm$ interfere with the growth of intricacy. Restricting the discussion to Case I with a mute channel, let one consider an intricon $a$ whose wave function $\varphi_a$ belongs to $\langle \rho_{A11} \rangle$. When it collides with a non-intricate atom $a'$, the density matrix $\rho_{a'}$ of $a'$ consists of three mutually incoherent parts, which are respectively partial traces on $\langle \rho_{A+\alpha} \rangle$, $\Omega_+$ and $\Omega_-$. When a state of $a'$ belongs to $\langle \rho_{A+\alpha} \rangle$, it interacts only with $a$ through its part belonging to $\langle \rho_{A11} \rangle$. The two-particle scattering matrix for the pair $aa'$ is governed by $H_A$ and unitary, so that $a'$ becomes intricate with $|1\rangle$ after the collision, increasing the number of intricons but conserving $p_1$.

When the state of $a'$ belongs to $\Omega_+$, there can be again a collision with $a$, but the two states of $a$ and $a'$ cannot be given definite phases and, although there is a scattering cross-section, there is no scattering amplitude (or rather its average on arbitrary phases is zero). There is then no constraint from unitarity and, when $a'$ becomes another intricon, it can bring also to $\langle \rho_{A11} \rangle$ an addition to its trace $p_1$. In the case of a state of $a'$ belonging to $\Omega_-$, it brings a negative probability to $\langle \rho_{A11} \rangle$. One saw, in the initial discussion of predecoherence in Section 2, that these two effects appear respectively as addition and subtraction to $\langle \rho_{A11} \rangle$, but the difference is now that they are no more simultaneous losses and reconstructions, because of saturation, In other words, $\Omega_+$ and $\Omega_-$ are independent whereas $\delta\rho_+$ and $\delta\rho_-$ were closely linked together and acted simultaneously. This simultaneity being damped by transport and mutual interactions, a collision of *a* with a state of *a'* in $\Omega$ occurs either with a state belonging to $\Omega_+$ or a state belonging to $\Omega_-$, with no precedence of $\Omega_-$ over $\Omega_+$. This property is obviously essential for a link of collapse with predecoherence, and it stands as a guess whose proof seems far from reach. Nevertheless, one will take it as an assumption and see whether it could have significant consequences.

A new question is then to decide whether fluctuations in ($p_1, p_2$) arise actually from incoherent fluctuations in the number of intricons. What is the relation between these quantities? This question appears nontrivial since, in the case when $p_1 = 1$, there would be fluctuations in the number of intricons, but there should be none in $p_1$. The same is true when $p_1 = 0$ and it suggests that $\langle \delta p_1^2 \rangle$ is probably proportional to the product $p_1 p_2$.

One can guess the desired relation between fluctuations in probabilities and in the number of intricons through a comparison between generic values of $p_1$ and the case $p_1 = 1$. Denoting by $n_1(x)$ the space density of intricons in the generic case and $n(x)$ when $p_1 = 1$, the Bethe-Bloch equation yields immediately the relation $n_1(x) = p_1 n(x)$, when there are only excitons and no secondary intricons. This proportionality implies also $n_1 = p_1 n$ for the total average number of excitons and $n_{1\beta} = p_1 n_\beta$ in a macroscopic cell $\beta$ inside the reactive region, because the history of intricacy in the two cases (averages being meant here as quantum averages and not predecoherent ones). But the eigenvalues of the number operator $N_\beta$ of excitons, with average value $n_\beta$, are integers and one can therefore consider that the excitons behave like particles, their space repartition being carried by their wave functions.

If one could apply the same considerations to intricons, one could label the intricons by a set $S$ of $n$ indices $\nu$ in the case $p_1 = 1$ and by a set $S'$ of $n_1$ indices in the generic case. The set $S'$ can be chosen as a subset of $S$ respecting approximately the space distribution. Let one consider then a collision of an arbitrary generic intricon with an atom in a state belonging to $\Omega_+$. The collision produces an increase $\delta n_1$. Because this intricon, whose label belongs to $S'$ corresponds also to an intricon with the same label in $S$, $n$ also increases by $\delta n_1$ from the same collision. The ratio $p_1 = n_1/n$ becomes then $p_1 + \delta p_1$, with

$$\delta p_1 = (n_1 + \delta n_1)/(n + \delta n_1) - n_1/n, \text{ or finally } \delta p_1 \approx p_2 \delta n_1 / n. \quad (3.10)$$

A remarkable property of this expression is to satisfy immediately the condition $\delta p_1 = 0$ when $p_1 = 1$.

When one considers the mute channel 2 in Case I, a change $\delta p_2 = -\delta p_1$ results necessarily from the fundamental property $p_1 + p_2 = 1$. One would like however to understand the mechanism producing this change, since Channel 2 is mute and ignores Channel 1 in principle. A consideration of the two previous sets $S$ and $S'$ of labesl in Channel 1 provides the necessary explanation as follows: The collisions of intricons with non-intricate atoms, labeled in the set $S - S'$, would have occurred for $p_1 = 1$ and they did not for a generic value of $p_1$. But the predecoherent matrices $\Omega_\pm$ ignore intricacy and their actions compensate each other on average on $\langle \rho_{A+\alpha} \rangle$, but there is no reason why this compensation should hold separately for $\langle \rho_{A11} \rangle$ and $\langle \rho_{A22} \rangle$. The fluctuations arising from intricons with label in $S - S'$ are not compensated in the generic case, and they appear as uncompensated transitions in Channel 2. Hence one can understand why $\delta p_2$ occurs, the reason of its sign and its explicit value.

One thus arrives at a rather clear overall pattern: There are collisions of atoms in the two channels, with or without secondary cascades bringing out records at a later stage. These events are independent and the respective developments in the two channels ignore each other. In other words, in Channel 1, the generation of excitons, development of intricacy, secondary cascades and so on will be identical whether $p_1 = 1$ (in which case the final outcome is certain) or when $p_1$ has another value. But whereas the events inside the two channels are independent, this is not so for their respective squared quantum amplitudes $p_1$ and $p_2$. The two channels share the predecoherent waves $\Omega_\pm$. There are interactions between copies of the channels and atomic states in these waves, which give rise to fluctuations in probabilities and a doubling of copies after a time $\delta t$. This is equivalent to a Pearle Brownian motion in probability space of two channels continuing to ignore each other in their inner dynamics.

*Note\**: The validity of (3.10) is questionable, because of its bold assimilation of intricons to a category of particles having its own number operator. One will use this formula however when evaluating fluctuations in probabilities. This is part of the guesswork, or of conjectures in the present work.

An attempt to check (3.10) is however in progress. It consists in considering that the Bethe-Bloch equation does not yield only an average value $n(x)$ for the density of excitons, but also gives in principle predictions for the squared standard deviations $\Delta n_\beta^2$ in cells $\beta$. One can introduce this "information" in (3.5) and draw the consequences for related deviations in $p_1$, but one encounters difficulties when going from addition in the exponent (3.5) to factorization in density matrices. Microlocal projection operators for a cell can be used for that purpose[20], but the calculations, which are long and tricky, have not been successfully completed. Rough estimates tend to confirm (3.10), but they remain questionable.

*Quantitative estimates*

To make an explicit calculation as short as possible, one considers Case II involving two tracks (generalization being straightforward). The two beams are identical with the same energy and differ only by the distant positions of the two tracks and their probabilities at time $0_+$. One considers only the initial stage of reduction when cascades have not yet developed. Attention is concentrated on excitons and on intricons resulting from them. Ions and electrons, which are in much smaller number, are also neglected. Previous notations are used, except for distinguishing the channels by an index 1 or 2. One has then $n = n_1 + n_2, n_1 = p_1 n, n_2 = p_2 n$ at a time $t$.

The locality of intricon-atom collisions is essential and, to take it into account, one covers each reactive region by a set of narrow cells with a width $\Lambda$, which are denoted by $\beta$ and transverse to the track. In the expression (3.4), one introduces approximate factorizations so that

$$\langle \rho_{Rjj} \rangle \approx p_j \prod_\beta \rho_{\beta j}, \qquad (3.11)$$

in which the density matrices of cells have trace 1.

Let the index $\beta$ now refer to a cell in Channel 1, $n_{\beta 1}$ being the average number of intricons in it and $n_\beta$ the corresponding reference number of intricons (which would occur in $\beta$ at time $t$ if $p_1$ were equal to 1. One has then $n_{\beta 1} = p_1 n_\beta$. Since the density matrices $\rho_{\beta 1}$

20have trace 1, the relation (3.10) between fluctuations in the number of intricons and fluctuations in probabilities during a time $\delta t$ becomes

$$\delta Tr(\rho_{\beta 1}) = p_2 \delta n_{\beta 1} / n_\beta. \tag{3.12}$$

Fluctuations in $n_{\beta 1}$ result from collisions of various intricons in $\beta$ with atoms whose state belongs either to $\Omega_+$ or $\Omega_-$. The corresponding changes in $n_{\beta 1}$ are respectively positive or negative and they cancel on average since $\Omega_+$ and $\Omega_-$ are closely similar locally. Letting then $\delta$ denote the probability for an individual intricon to collide with an atom with state in $\Omega_+$ during the time $\delta t$, the number of collisions arising from $\Omega_+$ or $\Omega_-$ is $n_{\beta 1}\delta$ and the squared standard deviation resulting from increase and decrease of $n_{\beta 1}$ by $\Omega_+$ and $\Omega_-$ is $\Delta^2 = n_{\beta 1}\delta$. Using (3.10), one gets then for the squared standard deviation arising from collisions in Channel 1

$$\langle (\delta Tr \rho_{\beta 1})^2 \rangle = (p_1 p_2^2 \delta)/n_\beta. \tag{3.13}$$

Bringing this into (3.11), one gets

$$\langle (\delta p_1)^2 \rangle = (p_1 p_2^2 \delta) \sum_\beta n_\beta^{-1} \approx (p_1 p_2^2 \delta) N(\beta)/n_\beta. \tag{3.14}$$

In the last expression, one did not try account for the evolution of slowing down along the track and introduced simply the number $N(\beta)$ of cells along the track.

But every fluctuation $\delta p_1$ implies a fluctuation $\delta p_2 = -\delta p_1$ and the relations

$$A \equiv \langle (\delta p_1)^2 \rangle = \langle (\delta p_2)^2 \rangle = -, \tag{3.15}$$

are quite general. There is therefore a fluctuation of $p_2$ in Channel 2 rising from collisions in Channel 1, still given by (3.14). The fluctuations resulting from collisions in both channels are therefore expressed with the notation in (3.14) by

$$A \approx \{p_1 p_2 (p_1 + p_2)\delta\} N(\beta)/n_\beta = (p_1 p_2 \delta) N(\beta)/n_\beta \tag{3.16}$$

Let one then make rough numerical estimates. Denoting again by $\tau$ the mean free time of collision for atoms, one has $\delta = \delta t/\tau$. If $L$ is the length of the track, one has $N(\beta) \approx L/\Lambda$. Considering the total number of intricons, soon after time $0_+$, as of the same order as the number of excitons, this is of order $E/e$, where $E$ is the energy of the alpha particle and $e$ the average energy of excitation of an exciton. The most poorly defined quantity is the minimal width $\Lambda$ of a cell, which will be taken as a few times the mean free path $\lambda$. Taking $\lambda = O(10^5 cm)$, $\tau = O(10^{-10} s)$, $E = 10$ MeV, $e = 10$ eV, $L = 10$ cm, one gets

$$\langle (\delta p_1)^2 \rangle = \langle (\delta p_2)^2 \rangle = -\langle \delta p_1 \delta p_1 \rangle = p_1 p_1 \delta t/\tau_c \tag{3.17}$$

where the time scale of collapse $\tau_c$ is of order $10^{-11}$s..

The ratio $L/\Lambda$ was probably overestimated and these numbers should not bring overconfidence in the results. One will add no further comments presently concerning possible generalizations,, what could happen later in cascades, possible fluctuations outside the reactive regions by photons arisng from the decay of excitons, other events in an outside electric circuit where the mean free time of electrons is significantly shorter , and so on. The main point is that, even if some guesswork was made, no basic obstructions was encountered against the idea that collapse could be due to a quantum mechanism. Nothing more will be claimed.

### 4. Other examples

The example of measurement in the previous sections was rather special and many extensions or discussions could be added. One will however avoid comments and mention only briefly two questions regarding many channels and the significance of mute channels.

In the case of Schrödinger's cat experiment, a radioactive source is part of the detector and can even stand inside it. Many possible different tracks behave then like so many measurement channels $j$ and carry some probabilities $p_j$. One can put nevertheless all of them together when discussing whether the radioactive source decayed or not. It is enough to proceed as in the previous section, except for using spherical cells $\beta$, centered on the source.

The question of many channels remains however and is tricky. The only simple property regarding them is a generalization of the formulas about correlations, which can be shown to become

$$\langle \delta p_j \delta p_{j'} \rangle = -p_j p_{j'} \delta t / \tau_c \quad \text{for } j \neq j', \tag{4.1}$$

$$\langle (\delta p_j)^2 \rangle = p_j (1 - p_j) \delta t / \tau_c. \tag{4.2}$$

The question asking how the different channels compete and how one of them wins finally the game remains anyway problematic.

Another question is again concerned with a paradigm of measurement, namely the Stern-Gerlach experiment: Two spin states of a spin-1/2 atom follow different trajectories along which two detectors are located. Why is it however that the two detectors are not independent and why, when there is a click in one detector, there is no simultaneous click in the other one?

The adaptation of the previous approach consists then in the necessary introduction of mute channels, implying to deal with *four* channels, namely: (1) The detector $D_1$ clicks and the other one $D_2$ does not. (2) $D_2$ clicks and $D_1$ does not. (3) Both detectors click. (4) Both detectors do not click. One finds then that mute channels, which had initially zero probability, are never created from predecoherence and never occur at the end. The connection of the two detectors, on the other hand, appears only through the property $p_1 + p_2 = 1$. This is clearly a manifestation of non-separability in quantum mechanics.

A similar question was raised by Nick Herbert and Bernard d'Espagnat. In that case, the question is directly concerned with non-separability, which is especially striking in the case of an EPR pair of photons when the two detectors have a space-like separation [16, 17]. This was considered in full detail in [13], Section 6, with the help of Bernard d'Espagnat. The calculations are rather lengthy, because the experiments are subtle, but they were given in that reference and no further comment will be made.

## 5. Assessments and perspectives

Some returns to principles are advisable before making conclusions, even in a sketchy way. Advisable also will be a use of the word "incoherence" here rather than "predecoherence". To begin with, one may point out that incoherence is not a one-way process going only from the environment to the apparatus. There is also an injection of incoherence from the apparatus towards the environment, as especially clear in high-energy physics where the environment of a detector is often made of other detectors. Incoherence is therefore a universal feature of physics, at least where macroscopic objects are present and have a classical behavior or a classical location. In that sense, the stochastic character of collisions from outside on a Geiger counter can be considered as a manifestation of external incoherence.

This standpoint does not contradict in principle the power and the value of idealization when, for instance, an atom is considered as an isolated system and studied for itself. The compatibility with a universe where incoherence dominates reality remains, because one can easily estimate in such a case the perturbations arising from the environment of the atom and check that they are negligible.

The difficulty –if there is one– is to rid one's mind of a tradition where one thinks primarily of quantum mechanics in terms of wave functions evolving unitarily, and takes this standpoint as so universal that, always, one could think that a non-isolated system belongs to a larger one that is isolated or behaves as if it were so. This kind of argument relies usually on the idea that a far-away part of the environment has no influence on a local system, because of the time necessary for its influence to be felt. But one saw here that an essential property of incoherence is its permanence, and the local incoherence acting at some time locally is always part of a permanent process forbidding a mental use of global unitary evolution. In that sense, one can never put a spatial limit to the environment.

Some physicists, well aware of this difficulty, tried to overcome it through recourse to a wave function $\Psi$ of the universe [25]. But the concept of $\Psi$ is rather vague and can receive many meanings. It has sometimes, in some papers not much more content than a Greek letter. Even if this $\Psi$ exists, it makes little sense from a mathematical standpoint. When one conceives a very wide Hilbert space and a version of quantum physics embedding space-time, it is obvious that $\Psi$ is extremely thorny. What are its properties of continuity? To which level in precision does one need to go down to make sure that it does not behave incoherently at some distance?

Although one may assume to have some grasp on it, for instance in string theory, $\Psi$ has certainly a high algorithmic complexity, or rather a high algorithmic depth (if one means by this depth that a computation of $\Psi$ on a Turing machine would become exponentially longer when a larger part of the universe is covered). Reversing the argument, one can then just as well conceive that $\Psi$ should be algorithmically random in the Kolmogorov-Chaitin sense [26]. This would be nice and the consistency of physics would become clearer: Algorithmic randomness would be responsible for the randomness of incoherence (predecoherence). Through Pearle's derivation of Born's rule or something like the present approach, one would understand why some squares of quantum amplitudes can become identical to the frequencies in a long series of measurements. One would be then allowed to call them "probabilities" without implying by this word that quantum mechanics is intrinsically probabilistic. Quantum mechanics would be only considered as so complex at a non-microscopic scale that unpredictability would be the essence of this probabilism, not because of ignorance as Laplace thought, but because of intrinsic inaccessibility.

There could also be still wider perspectives: The present work started rather long ago from an impression that the quantum principles are so deep that they could contain in germ every concept entering in their interpretation. This idea came out from some advances in interpretation, such as the removal of many paradoxes by the logic of consistent histories [27, 28, 20], the agreement of classical determinism with quantum dynamics [20], and last but no least, the concept and observation of decoherence [6-11]. All these results came directly from the quantum principles without any change in these principles. If the uniqueness of macroscopic reality could become also a consequence of the quantum principles, one would be not far from a deductive interpretation of quantum mechanics. One could also propose a new answer to the metaphysical problem of Reality [29], in which an absolute reality $\Psi$ (or a fine-grained history of the universe [30]) implies dynamically, through its own complexity, the uniqueness of empirical reality at a macroscopic scale.

This is of course extremely speculative however and one acknowledged on the other hand that the present work contains more guess than proof. Many errors occurrred during quite a few years when this research was made; but may be worth mentioning that, after the main lineaments had been found, the theory took shape by itself and led with no great difficulty to suggestive answers for the questions it raised.

As a conclusion, one must recognize again that guesswork has been exercised at some key point concerning the relation between predecoherence and collapse, where a thorough analysis and rigorous proofs would be needed. Nevertheless, I dare say that a fundamental consistency within quantum mechanics, encompassing measurement, appears more attractive than most proposals for the deep problem of objectification. Further investigations would be worth more thorough attempts.

## References

[1] J.A Wheeler, W.H. Zurek, *Quantum mechanics and measurement*, Princeton University Press (1983)
[2] ] E.P. Wigner, *Interpretation of quantum mechanics*, and *The problem of measurement*, Am. J. Phys. **31**, 6 (1963), reprinted in [1].
[3] J. von Neumann, *Mathematische Grundlagen der Quantenmechanik*, Springer, Berlin (1932). English translation by R. T. Beyer, *Mathematical Foundations of Quantum Mechanics*, Princeton University Press (1955)
[4] E. Schrödinger, *Naturwissenschaften* , **23**, 807, 823, 844 (1935), reprinted in [1].
[5] E. Borel, *Le hasard*, Alcan, Paris (1914)
[7] K. Hepp, E. H. Lieb, *Helvet. Phys. Acta,* **46**, 573 (1974)
[8] W.H. Zurek, *Phys. Rev. D* **26**, 1862 (1982)
[9] A.O. Caldeira, A. J. Leggett, Physica **A 121**, 587 (1983)
[10] E. Joos, H.D. Zeh, C. Kiefer, D. Giulini, K. Kupsch, I.O. Stamatescu, *Decoherence and the Appearance of a Classical World in Quantum Theory*, Springer, Berlin (2003)
[11] M. Brune, E. Hagley, J. Dreyer, X. Maître, A. Maali, C. Wunderlich, J.M. Raimond, S. Haroche, *Phys. Rev. Lett*, **77**, 4887 (1996)
[12] S. Haroche, J.M. Raimond, *Exploring the quantum*, Oxford University Press (2006)
[13] R. Omnès, Found. Phys., **41**, 1857 (2011)
[14] P. Pearle., Phys. Rev. **D 13**, 857. (1976), J. Theor. Phys. **48**, 489 (1979) ; Phys. Rev. **A 39**, 2277 (1989), arxiv.org. 0611211 and 0611212
[15] N.F.Mott, *Roy. Soc. Proc.* **A 126**, 79 (1929)
[16] A. Aspect, P. Grangier, G. Roger, *Phys. Rev. Lett.*, **49**, 91 (1982); A. Aspect, PJ. Dalibard, G. Roger, *Phys. Rev. Lett.*, **49**, 1804 (1982)
[17] A. Stepanov, H. Zbinden, N. Gisin, A. Suarez, *Phys. Rev. Letters*, **88**, 120404 (2002) ; *Phys. Rev. A*, 042115 (2003).
[18] L.S. Brown, *Quantum field theory*, Cambridge University Press (1992)

[19] L.D. Landau, E.M. Lischitz, *Statistical Physics*, Pergamon, London (1969)
[20] R. Omnès, *Interpretation of quantum mechanics*, Princeton University Press (1994)
[21] F. Laloë, private communication
[22] R. Balian, *From microphysics to macrophysics*, Springer, Berlin (2007)
[23] C. Caso et al. *The Review of Particle Physics*, The European Physical Journal C3 (1998)
[24] C. Kittel, *Introduction to Solid State Physics*, Wiley, New York (2005)
[25] A. Bassi, G.C. Ghirardi, Phys. Lett. **A275**, 373 (2000)
[26] G. J. Chaitin, *Randomness and Incompleteness*, World Scientific, River Edge, N.J. (1990)
[27] R.B. Griffiths, J. Stat. Phys. **36**, 219 (1984). *Consistent Quantum Mechanics*, Cambridge University Press, Cambridge (UK), (2002)
[28] M. Gell-Mann, J.B. Hartle, *Quantum mechanics in the light of quantum cosmology* in *Complexity, entropy and the physics of information*, SFI Studies in the science of complexity, Vol. VIII, ed. by W. Zurek, Addison-Wesley, Reading, MA (1990); *Classical equations for quantum systems*, Phys. Rev. **D47**, 3345 (1993)
[29] B. d'Espagnat, *On physics and philosophy*, Princeton University Press (2006)
[30] M. Gell-Mann, J.B. Hartle, *Decoherent histories quantum mechanics with one "real" fine-grained history*, arxiv.org. 1106. 0767